\documentclass{article}
\usepackage[english]{babel}
\usepackage[latin1]{inputenc}

\usepackage{spconf,amsmath,graphicx}
\usepackage{amsthm}
\usepackage{amssymb}
\usepackage{slashbox}
\usepackage{bbm}
\usepackage{array}

\theoremstyle{plain}

\newtheorem{prp}{Proposition}

\theoremstyle{definition}
\newtheorem{defn}{Definition}

\theoremstyle{remark}

\usepackage{subfigure}

\title{Track selection in Multifunction Radars for Multi-target \\tracking: an Anti-Coordination game}
\name{ Nikola Bogdanovi\'{c}, Hans Driessen, Alexander Yarovoy 
	\thanks{The work reported in this paper has been conducted as part of the Sensor Technology Applied in Reconfigurable systems for sustainable Security (STARS) project, see the website www.starsproject.nl; the multi-site tracking algorithm that is used in this paper has been developed under the SOS project funded by the European Commission (FP7 GA no. 286105).}}

 \address{Microwave Sensing, Signals and Systems\\ Delft University of Technology, The Nederlands\\
 E-mails: \{N.Bogdanovic, J.N.Driessen, A.Yarovoy\}@tudelft.nl}

\begin{document}
\ninept
\maketitle
\begin{abstract}

In this paper, a track selection problem for multi-target tracking in a multifunction radar network is studied using the concepts from game theory. The problem is formulated as a non-cooperative game, and specifically as an anti-coordination game, where each player aims to differ from what other players do. 
The players' utilities are modeled using a proper tracking accuracy criterion and, under different assumptions on the structure of these utilities, the corresponding Nash equilibria are characterized. 
To find an equilibrium, a distributed algorithm based on the best-response dynamics is proposed. 
Finally, computer simulations are carried out to verify the effectiveness of the proposed algorithm in a multi-target tracking scenario.

\end{abstract}
%


\section{Introduction}

Radar networks that employ multiple, distributed stations 
offer significant advantages over 
standalone radars, 
in terms of providing diversities and enhancing tracking and detection performance.
Furthermore, 
recent advances in sensor technologies enabled a large number of controllable degrees of freedom in modern radars. 
One such system is the 
Multifunction Radar (MFR), which employs an electronically scanned phased array composed of individually controlled radiating elements~\cite{hero2011sensor}\nocite{sabatini_tarantino_1994}-\cite{richards_scheer_holm_2010}.
%
Due to its beam and waveform agility, the MFR 
is capable to track multiple targets and perform new target search in the sector.
%
Thus, the MFR is much more flexible than conventional, dedicated radars 
by being capable of performing different functions - volume surveillance, weapon control, and multiple target tracking to name a few.
%
In this paper, we focus on the latter function~\cite{blackman1999design}-\cite{Waveform_agile_4775880}; specifically, each MFR radar aims at tracking several targets. 

Even for a standalone 
MFR, the radar resource management plays a crucial role so as 
to efficiently allocate resources to achieve specified objectives while conforming to operational and technical constraints~\cite{Ding_4564804}, \cite{katsilieris2015sensor}.
%
%
Most of the existing approaches to MFR radar resource management 
roughly fit into the following two categories~\cite{charlish2011autonomous},~\cite{narykov2013algorithm}. 
%
The first category consists of the rule-based techniques~\cite{van1993phased}\nocite{koch1999adaptive}-\cite{coetzee2005multifunction}, which control the resource allocation parameters indirectly, under low computational burden
. However, it is hard to say what performance 
can be achieved since it highly depends on the application scenario and on the sensors being deployed. 
%
The other category is related to the methods that formulate the problem as an optimization one; and thus, they may achieve the optimal performance, 
 see~\cite{Hans_1591902}\nocite{hansen2006resource}-\cite{Djonin_4915772},~\cite{hero2011sensor},~\cite{Waveform_agile_4775880} and the references therein.

Note that, in the network setting, the first category of approaches is difficult to be extended, while the second one may involve excessive complexity due to the network dimension.   
To reduce 
such complexity, 
in this work we propose a distributed approach based on game theory 
so as to 
model track selection for multi-target tracking in an MFR network.

Game theory is the mathematical study of conflict and cooperation between intelligent rational decision-makers~\cite{shoham2008multiagent}. 
%
In addition to its traditional research areas such as economics and political sciences, 
%
over the last decade  game theory (GT) is being applied to 
signal processing and wireless communications. 
%
This is mainly due to the issues dealing with (distributed) networking~\cite{Han_2012_book}, \cite{bacci_2015game}, 
such as 
power control~\cite{Saraydar_983324}, 
antennas' beamforming~\cite{Larsson_4604732}, multiple input multiple output (MIMO) communications~\cite{scutari2008competitive}, 
channel allocation~\cite{ch1_j_Feleg4907463}, 
adaptive estimation~\cite{Yu2013a},~\cite{GT_icassp2015}, to name a few.
%
%
%
%
%
%
More recently, GT has been applied to solve certain radar problems,  mostly related to the 
MIMO radar networks.
%
For instance, the problem of waveform design has been tackled;  in~\cite{g_2012_game} 
by formulating a two player zero-sum (TPZS) game 
between the radar design engineer and an opponent, 
and in~\cite{code_design_piezzo2013non} by a potential game 
in which the radars choose among the pre-fixed transmit 
codes. 
%
Next, the interaction between a smart target and a MIMO radar is modeled as a TPZS 
game~\cite{Jammer_6025317}, where the mutual information criterion was used 
in the utility functions. 
%
Also, the problem of transmission power 
management 
was addressed in~\cite{Bacci_6250454}, 
 \cite{Chen_coop_game_7104065}. 
Initially, the power control problem was formulated in~\cite{Bacci_6250454} assuming the presence of some interference due to the other radars' transmissions. A non-cooperative game was used for modeling and a distributed algorithm converging to a Nash equilibrium was proposed.
On the other hand, in reference~\cite{Chen_coop_game_7104065}, 
a coalitional game theoretic solution concept called the Shapley value was employed to distribute 
a given power budget among all transmitting radars.  
%
More related to the application scenario in this article, 
the work in~\cite{charlish2012multi} utilize a market mechanism, called the continuous double auction, in order to choose the global optimum parameters for each individual task given the global (finite) resource constraint.


\begin{figure}
	\centering
	\includegraphics[width=0.575\linewidth]{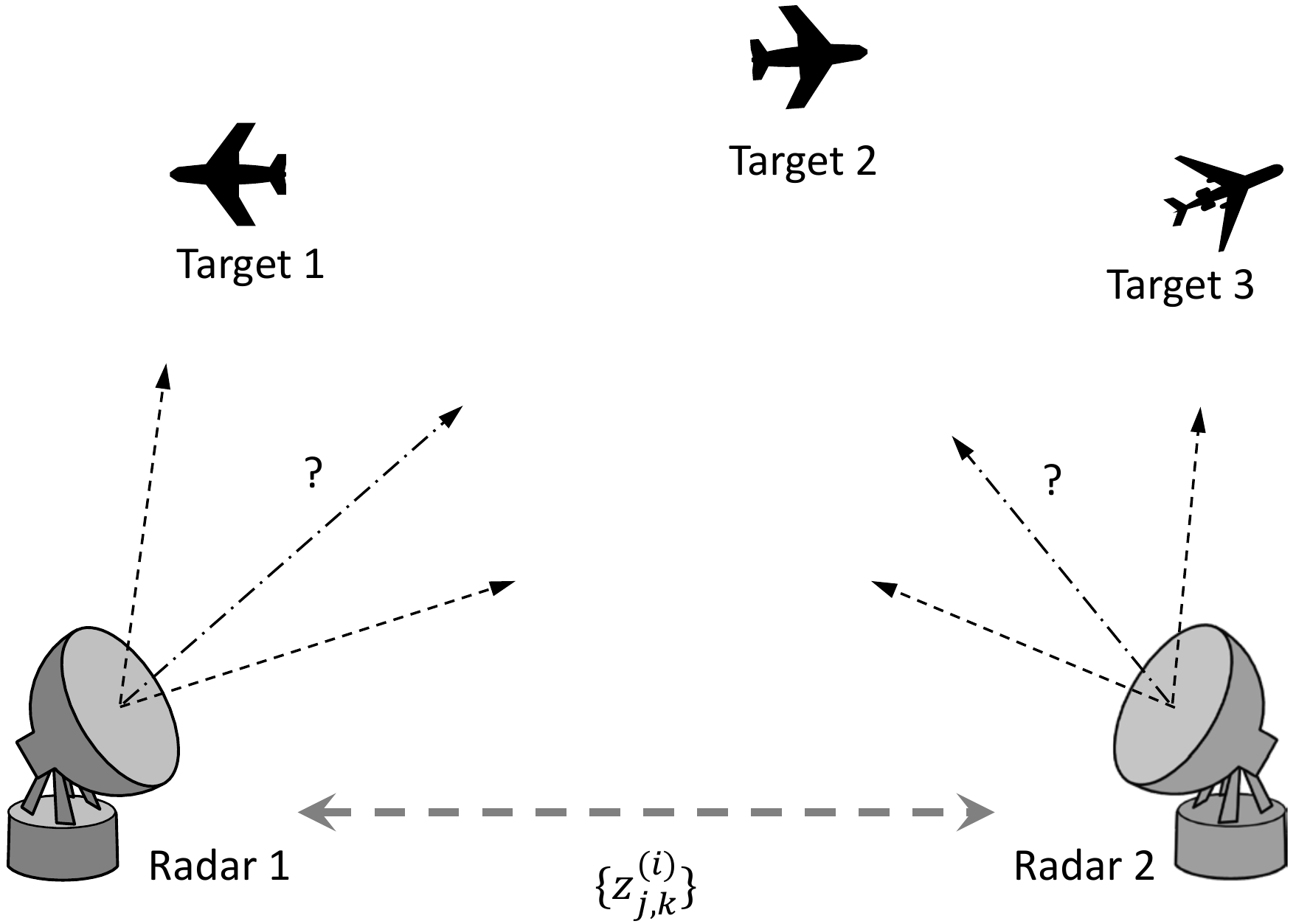}
	\caption{A track selection problem in multi-target tracking.}
	\label{fig:radar_network_1_1}
\end{figure}

In contrast to the aforementioned literature, 
in this 
paper we formulate 
a new problem of track selection 
for a multi-target tracking scenario in a resource-limited MFR network using the non-cooperative games, the dominant branch of GT~\cite{shoham2008multiagent}. An example of such a scenario is depicted in Fig.~\ref{fig:radar_network_1_1}.
To the best of our knowledge, this is the first non-cooperative GT contribution dealing with multi-target tracking. 
%
%
	Due to its nature, the problem is modeled as a coordination game which is known to have several Nash equilibria.
	Then, the equilibria are 
	characterized in terms of their existence conditions and efficiency.
	Finally, to find an equilibrium, a distributed algorithm based on the best-response dynamics is proposed and its 
	effectiveness for the track selection issue in multi-target tracking is demonstrated.
%
%

\section{Problem formulation}
\label{Sec_3_System}

Suppose that there are multiple MFR 
 radars and several targets to be tracked 
whose number is known exactly 
and 
their current positions 
approximately
, see Fig.~\ref{fig:radar_network_1_1}. We denote the set of radars by $\mathcal{N}$, while the set of targets is denoted by $\mathcal{T}$. The targets are assumed to be well-separated; thus there is no data association problem and different transmission beams are required so as to illuminate the targets. 
Consider that there is no fusion center 
and that each radar aims at tracking all targets simultaneously. Next, each target $j \in \mathcal{T}$, at each discrete time $k$, follows the so-called white noise constant velocity model~\cite{blackman1999design},~\cite{Hans_1591902} given by
\begin{align} 
 \mathrm{x}_{j,k}  &= F\cdot \mathrm{x}_{j, k-1} + w_{j, k-1} \\ 
z_{j,k}^{(i)} &=  h^{(i)}_j(\mathrm{x}_{j,k}) + \nu^{(i)}_{j,k}
\end{align}
where the state vector $ \mathrm{x}$ for each target $j$ is comprised of the two dimensional coordinates and velocity, i.e., $\mathrm{x}_j=[ x_j ,y_j , v_{j,x}, v_{j,y}]^T$, while $F$ is a $4\times4$ matrix corresponding to the deterministic target dynamics given as 
$F=\left[\begin{smallmatrix} 
1 & t_u \\
0 & 1
\end{smallmatrix}\right] \otimes I_2$, with $\otimes$ being the Kronecker product, $I_2$ stands for a $2\times2$ identity matrix and $t_u$ is the update time that is fixed. The process noise $w$ is Gaussian with zero mean and covariance $Q=\sigma^2_w \cdot \left[\begin{smallmatrix} 
t_u^3/3 & t_u^2/2 \\
t_u^2/2 & t_u
\end{smallmatrix}\right] \otimes I_2$, where $\sigma^2_w$ models maneuverability. At each radar $i \in \mathcal{N}$, the measurement vector $z_{j,k}^{(i)}$ consists of range and azimuth, i.e., $z_{j,k}^{(i)}=\left[r_{j,k}^{(i)}  , a_{j,k}^{(i)}\right]^T$, while the nonlinear transformation  $h^{(i)}_j(\mathrm{x}_{j})$ is given by $$ h^{(i)}_j(\mathrm{x}_{j})=\begin{bmatrix} 
\sqrt{(x_j-x_i)^2 + (y_j-y_i)^2} \\
 \mathrm{arctan}((y_j-y_i)/(x_j-x_i)) \end{bmatrix}.$$
The coordinates $(x_i,y_i)$ of each radar $i \in \mathcal{N}$ are assumed known. Finally, the measurement noise $\nu^{(i)}_{j}$ is zero-mean Gaussian with covariance $R_{j,i}=\mathrm{diag}\left\{[\sigma^{(i)}_{r_j}]^2,[\sigma^{(i)}_{a_j}]^2\right\}$.


The radars have limited time budget in sense that they cannot take measurements of %
all targets 
during the same time slot. 
The number of measurements per scan that each radar can make is given by $m < \vert\mathcal{T}\vert$. 
Since there is no central entity that may coordinate actions of the radars, 
a distributed solution is needed. %
The interaction among the radars is existing but limited to sharing, e.g., by broadcasting, 
the measurements $\{z_{j,k}^{(i)}\}$ 
related to the  previously selected targets. 
The number of transmissions each target $j$ is tracked by at one time slot is denoted as 
$m^t_j$. %
For notational simplicity, in the rest of this section we drop the index $j$ for targets.

At each radar $i$ and for each target $j$, the tracking process is performed by 
an Extended-Kalman Filter (EKF). Firstly, the prediction step occurs, i.e.,
%
\begin{align}
\mathrm{x}_{k\vert k-1}  &= F\cdot \mathrm{x}_{k-1\vert k-1}  \label{eq_predict_1}\\ 
P_{k\vert k-1} &=  F P_{k-1\vert k-1} F^T + Q  \label{eq_predict_2}
\end{align}
where $\mathrm{x}_{k\vert k-1}$ and $P_{ k\vert k-1}$ are the state estimate and the error covariance matrix for time step $k$ given all measurements till time step $k-1$. 
Then, the updating step takes place where each available measurement for target $j$ of some radar $n\in \mathcal{N}$ 
is used in a cyclic manner. In particular, for  each $p\in \{1,\ldots, m_j^t\}$, 
\begin{align}
K_{k}^{(p)}  &= P_{k\vert k}^{(p-1)} [H^{(p)}_{k,n}]^T \left(H^{(p)}_{k,n} P_{k\vert k}^{(p-1)} [H^{(p)}_{k,n}]^T  +R_{n}\right)^{-1}   \label{eq_update_1}\\ 
\mathrm{x}_{k\vert k}^{(p)}  &= \mathrm{x}_{k\vert k}^{(p-1)} +K_{k}^{(p)} \left( z_{k}^{(n)} - h^{(n)}\left(\mathrm{x}_{k\vert k}^{(p-1)}\right)\right) \label{eq_update_2}\\ 
P_{k\vert k}^{(p)}  &=  \left(I- K_{k}^{(p)} H^{(p)}_{k,n}\right) P_{k\vert k}^{(p-1)} \label{eq_update_3}
\end{align}
where $P_{k\vert k}^{(p)}$ denotes the error covariance matrix after $p$ incremental updates at the same time step $k$, with 
$P_{k\vert k}^{(0)}=P_{k\vert k-1}$ and $\mathrm{x}_{k\vert k}^{(0)}=\mathrm{x}_{k\vert k-1}$.
%
%
The linearized measurement matrix of radar $n$ at time $k$ is $H^{(p)}_{k,n}= \partial h^{(n)}/ \partial \mathrm{x}$ evaluated at $\mathrm{x}_{k\vert k}^{(p-1)}$.
Note that, due to the fact that the 
position of each radar is known, the radars do not need to exchange $\{H_{k,n}\}$ matrices in order to implement the algorithm above.

	In the following, we study a natural game theoretic variant of this problem. Specifically, we assume that the radars are 
autonomous decision-makers interested in optimizing their own tracking performance. 
%
%
	The fact that each radar autonomously and rationally decides to track the targets that increase its utility can be modeled as a 
one-stage non-cooperative game in normal form.

\section{Game-theoretic model}
\label{Sec_4_GT_model}


Here, we formulate the track selection problem in multi-target tracking as a non-cooperative game in normal form, which is the most fundamental representation type in game theory~\cite{shoham2008multiagent}. 
Note that there are many classes of normal-form games; however, in this work we focus on coordination games, which do not rest solely upon conflict among players. Instead, as their name suggests, more emphasis is put on the coordination issue where players may have an incentive to conform with or to differ from what others do. In the latter case, this kind of games are usually called \textit{anti-coordination games}~\cite{shoham2008multiagent}, \cite{nisan2007algorithmic}-\cite{bramoulle2007anti}.

We assume that the players are rational and their objective is to maximize their payoff, i.e., the tracking accuracy of all targets. %
Formally, the 
track selection game $(\mathcal{N}, \mathcal{S}, u)$ has 
the 
subsequent components:
\begin{itemize}
	\item The players are the radars represented by the set $\mathcal{N}$.
	\item The strategy of each radar $i$ is represented by a $\vert\mathcal{T}\vert$-tuple $s_i =(s_{i,1}, s_{i,2}, \ldots , s_{i,\vert\mathcal{T}\vert})$ where $s_{i,j} =a$ 
	if radar $i$ devotes $a$ transmission beams to a target $j$, with $a\leq m$. Each strategy-tuple has at most $m$ transmissions, i.e., $\sum_{j=1}^{\vert\mathcal{T}\vert} s_{i,j} \leq m$. 
Also, note that $m^t_j=\sum_{i=1}^{\vert\mathcal{N}\vert} s_{i,j}$. 
	%
Each vector $s = (s_1, \ldots , s_N) \in \mathcal{S}$ is called an action or strategy profile, and $s_{-i} = (s_1, \ldots , s_{i-1}, s_{i+1}, \ldots , s_N)$ is defined as a strategy
	profile $s$ without player $i$'s strategy.
	\item The utility for each radar $i$ is given by
	\begin{gather}\label{eps:GT_model_eq1}
		\begin{split}
			u_i (s_i, s_{-i}) =     \sum_{j=1}^{\vert\mathcal{T}\vert} \left( \mathrm{gain}_j(m^t_j)  - c \cdot \tau_j \right),
		\end{split}
	\end{gather}
	where the term 
		$\mathrm{gain}_j(m^t_j)$ represents the  tracking accuracy gain for target $j \in \mathcal{T}$ and it is defined by
		\begin{equation}\label{eps:gain_eq}
			\mathrm{gain}_j (m^t_j) =\begin{cases}
				\mathrm{Tr} \left\{ P_{j, k\vert k-1}  - P_{j, k\vert k}^{(m^t_j)} \right\},&\text{if} \,\, m^t_j \geq 1 \\
				0, & \text{otherwise}
			\end{cases}
		\end{equation}
where  all radars are assumed to have the same initial guesses $\mathrm{x}_{j, 0\vert 0}$ and $P_{j, 0\vert 0}$.
Finally, the normalization coefficient $c$ indicates delay importance, while 
		$\tau_j$ is the delay function given as
		\begin{equation}
			\tau_j=\begin{cases}
				0, & \text{if} \,\,\, m^t_j \geq 1 \\
				1, & \text{otherwise}
			\end{cases} \quad .
		\end{equation}
		
\end{itemize}

In other words, the strategy of radar $i$ defines the number of transmissions per each target, at a given time slot.
Due to the fact that radars share their measurements
, their tracking accuracy gains for a specific target are 
dependent on all radars' measurements related to that target. 
%
%
If set to be non-negative, the delay importance coefficient can serve as a mechanism for punishing radars in case where not all targets are being tracked. Otherwise, if set negative enough, it gives incentives to radars to focus all their resources on less targets. In our study, it is of interest to study the former case.  



\begin{figure}[!t]
	\centering 
	\subfigure[]{
		\includegraphics[width=0.22\textwidth]{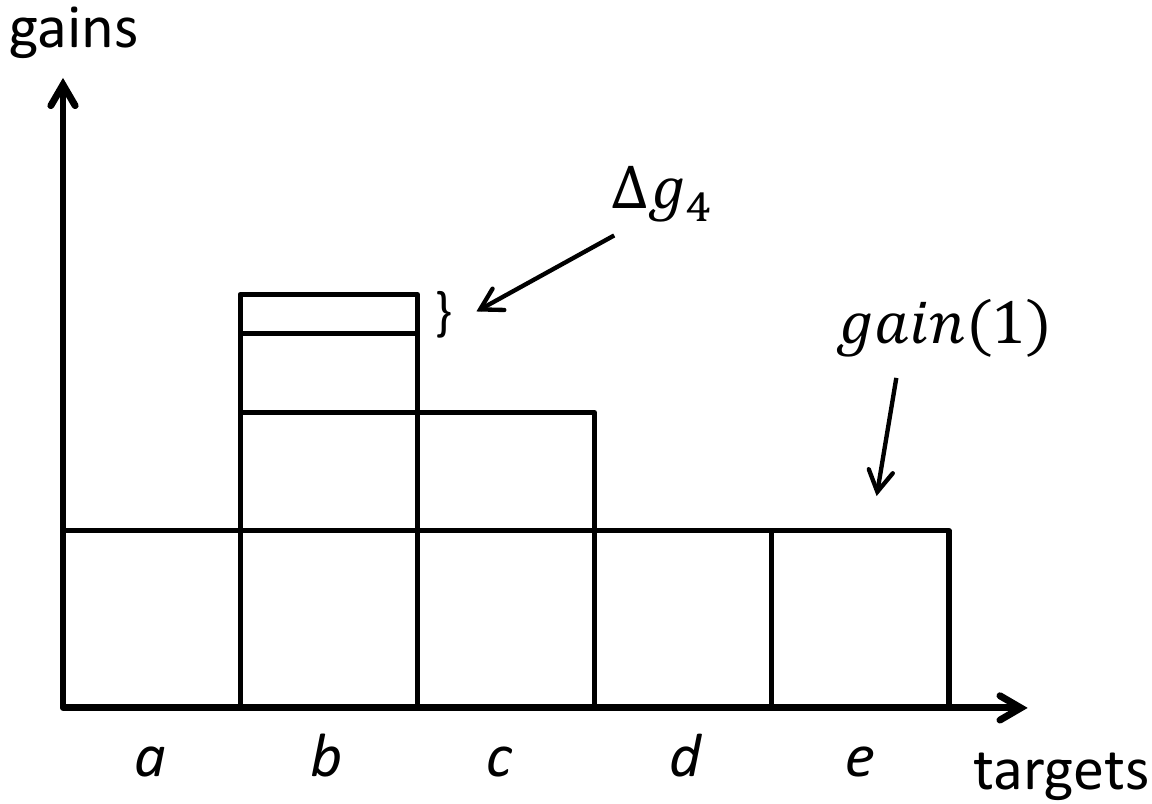}
		\label{fig:example1a}
	} \hspace{0.1cm}
	\subfigure[]{
		\includegraphics[width=0.22\textwidth]{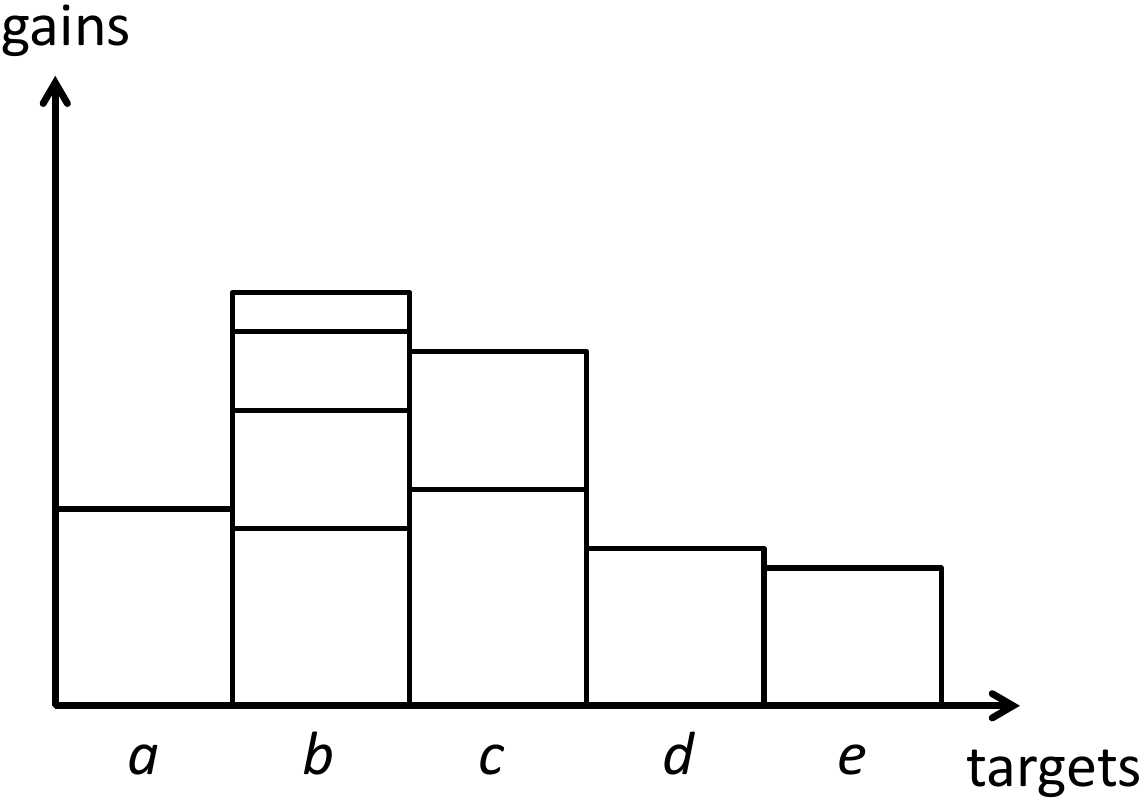}
		\label{fig:example1b}
	}
	\caption{An 
example of a 
track allocation where 
		$\mathcal{T}=\{a, \ldots, e\}$, and $\vert\mathcal{N}\vert=m=3$. Each box represents a gain increment due to a measurement, and the number of measurements per target, $m^t_j$, varies between 1 and 4 across targets in $\mathcal{T}$
		. In case~\subref{fig:example1a}, the gains are equal for the same number of measurements, while in case~\subref{fig:example1b} they differ. 
	}  
	\label{fig:example1_1}
\end{figure}

In practice, the gain function in~\eqref{eps:gain_eq} can be assumed to be increasing in the number of measurements, at least in mean sense. 
%
Note that the gain in~\eqref{eps:gain_eq} can be expressed as 
$ \mathrm{gain}_j\left(m^t_j\right)= \sum_{p=1}^{m^t_j}  \Delta g^{(j)}_{p}$, where $\Delta g^{(j)}_{p}=\mathrm{Tr}\{ P_{j, k\vert k}^{(p-1)}  - P_{j, k\vert k}^{(p)} \}$ and $\Delta g^{(j)}_{1}=\mathrm{gain}_j(1)$. Also, 
in a real system it reasonable to assume that an estimation accuracy gain increment $\Delta g^{(j)}_{p}$ decreases as the order of measurements $p$ grows, i.e.,  $\Delta g^{(j)}_{p} > \Delta g^{(j)}_{p+1}$. For the analysis in the sequel, the following two cases are distinguished:
\begin{itemize}
\item[a)] $\Delta g^{(j)}_{p}=\Delta g_{p}$,  
for all $j \in \mathcal{T}$ and $p \in \{1, \ldots, m^t_j\}$
\item[b)] $\Delta g^{(j)}_{p} \neq \Delta g^{(\ell)}_{p}$, 
for $j \neq \ell$, 
and $\mathrm{min}_{j \in \mathcal{T}}\Delta g^{(j)}_{p} > \break \mathrm{max}_{j \in \mathcal{T}} \Delta g^{(j)}_{p+1}$.
\end{itemize}
Case a) represents an idealistic case where all nodes would have very similar measurements among themselves and related to all targets, see Fig~\ref{fig:example1_1}\subref{fig:example1a}. A more realistic scenario, corresponding to case b), is illustrated in Fig~\ref{fig:example1_1}\subref{fig:example1b}. In the following section, the track selection game for both cases will be analyzed.

\section{Nash equilibria}
\label{Sec_Nash}

In general, one can reason about multiplayer games using solution concepts, i.e., principles according to which interesting outcomes of a game can be identified. 
Although there are many solution concepts in the game-theory literature, a basic and the most widely accepted one is the Nash equilibrium. %
Formally, in case in which players make deterministic choices (pure strategies)
the Nash equilibrium is defined as follows%
~\cite{shoham2008multiagent}.

\begin{defn}
A strategy profile $s = (s_1, \ldots , s_N)$ is a pure-strategy Nash
	equilibrium (NE) if, for all players $i$  and for all strategies $s^{\prime}_i \neq s_i$, it holds that 
$u_i (s_i, s_{-i}) \geq  u_i (s^{\prime}_i, s_{-i})$. 
\end{defn}
In other words, in an NE, no player can unilaterally improve its utility by taking a different strategy. Also, it is important to define Pareto domination and Pareto optimality.

\begin{defn} 
	Strategy profile $s$ Pareto dominates strategy profile $s^{\prime}$ if  $\forall i \in \mathcal{N}$, $u_i(s) \geq u_i(s^{\prime})$, and there exists some $n \in \mathcal{N}$ for which $u_n(s) >  u_n(s^{\prime})$. Also, strategy profile $s$ is Pareto optimal (PO) 
	if there does not exist another strategy profile $s^{\prime} \in \mathcal{S}$ that Pareto dominates $s$.
\end{defn}

Generally, in a coordination game, there are multiple NE. 
If the players have the same payoffs
, and the equilibria are equal, the game 
is a \textit{pure} coordination one.
In fact, in such a game, all NE are PO. On the other hand, in a  \textit{ranked} one, 
the NE differ and usually there is only one PO equilibrium~\cite{book_rasmusen_2001}.

%
%

Now, the main findings related to the NE for cases (a) and (b) are provided.

\begin{prp}
\label{b_slucaj}
For case (a), any track assignment is a PO NE, if 
$c \geq 0$ and $\sum_{j=1}^{\vert\mathcal{T}\vert} s_{i,j} = m$, and if
\begin{itemize}
\item  $m^t_j \leq 1$, $\forall j \in \mathcal{T}$, for a scenario where $\vert\mathcal{N}\vert \cdot m \leq \vert\mathcal{T}\vert$
\item  $\mathrm{max}_{j, \ell \in \mathcal{T}} \{ \vert m^t_j - m^t_{\ell}\vert \} \leq 1$, $\forall j,\ell \in \mathcal{T}$, for a scenario where $\vert\mathcal{N}\vert \cdot m > \vert\mathcal{T}\vert$.
\end{itemize} 
\end{prp}

Firstly, 
let us assume 
	that there is a radar $i$ such that $\sum_{j=1}^{\vert\mathcal{T}\vert} s_{i,j} < m$ and that the corresponding $s^{\ast}$ is an NE. 
	Then, radar $i$ can change 
its strategy by taking an additional measurement. 
	Due to the fact that  
	the radar's gain function in~\eqref{eps:gain_eq} is increasing in the number of measurements, its utility will be increased. 
	But that contradicts our initial assumption that  $s^{\ast}$ is an NE; thus, 
%
%
%
as per our intuition, each radar should make all possible transmissions toward the target(s) at each time instant. 
%
%
Next, 
note that if the total number of measurements is less than or equal to the number of targets, the condition related to $c$  ensures 
that the radars are punished if more than one measurement in total is devoted to the same target. 
Also, due to the structure of gain function, NE are precisely $\frac{\vert\mathcal{T}\vert!}{\left(\vert\mathcal{T}\vert -\vert\mathcal{N}\vert \cdot m\right)!}$ outcomes in which each measurement is devoted to a distinct target.
On the other hand, if $\vert\mathcal{N}\vert \cdot m > \vert\mathcal{T}\vert$, condition $c \geq 0$ promotes all targets to be covered. Here, each NE corresponds to a balanced allocation. For instance, the allocation in Fig~\ref{fig:example1_1}\subref{fig:example1a} is not an NE since the payoffs can be increased if some player moves its measurement from target $b$ to any other target. 
Finally, %
since the gain of any target is the same for the same number of measurements,  
%
the game appears to be 
a pure anti-coordination one. 
%
%
%
%
%
Thus, every NE is also Pareto optimal.

\begin{prp}
	\label{less_case_c}
	For case (b), any track assignment is an NE, if 
$c \geq 0$ and $\sum_{j=1}^{\vert\mathcal{T}\vert} s_{i,j} = m$, and if,
	\begin{itemize}
		\item  
		for a scenario with $\vert\mathcal{N}\vert \cdot m \leq \vert\mathcal{T}\vert$,
		each radar chooses its 
		most accurate target that has not been selected,
		\item for 
$\vert\mathcal{N}\vert \cdot m > \vert\mathcal{T}\vert$,
		the first $\left \lceil{\frac{\vert\mathcal{N}\vert \cdot m}{\vert\mathcal{T}\vert}}\right \rceil-1$ levels are filled in, i.e., $m_{j}^t \geq \left \lceil{\frac{\vert\mathcal{N}\vert \cdot m}{\vert\mathcal{T}\vert}}\right \rceil-1$,  $\forall j \in \mathcal{T}$,  and for the $\left \lceil{\frac{\vert\mathcal{N}\vert \cdot m}{\vert\mathcal{T}\vert}}\right \rceil$-th level each radar chooses its 
		most accurate target that has not been selected by others, where $\lceil{\cdot}\rceil$ is the ceiling function.
	\end{itemize} 
\end{prp}

Similar arguments hold as for Prop.~\ref{b_slucaj}. 
Yet, 
the game above seems to be a 
ranked anti-coordination game.
Note that here there are still multiple NE, but not all NE are necessarily equal, and hence, not every NE is Pareto optimal (only one is). So, the conditions above are not sufficient to have also a Pareto optimal NE.


To conclude this section, we provide a simple, distributed algorithm, based on the best response dynamics~\cite{shoham2008multiagent},
\cite{Han_2012_book}, to achieve an NE.
In the games above where $\vert\mathcal{N}\vert \cdot m > \vert\mathcal{T}\vert$, in general, two types of NE may arise,  
one where a radar illuminates only different targets and the other where it chooses the same target more than once.
In practice, it is of interest to exploit the radars' diversity; thus, we focus on the former type.
%
%
%
Let $\mathcal{T}_i$ denote the set of targets selected by radar $i$.
Then, for any initialization, at each time instant each radar $i \in \mathcal{N}$ performs the following steps:
\begin{itemize}
\item Count $m_j^t$, $\forall j \in \mathcal{T}$, and reallocate the measurements for $\forall j \in \mathcal{T}_i$ 
satisfying $s_{i,j}>1$ to a target $\mathrm{ arg min}_{\ell \in \{\mathcal{T}/\mathcal{T}_i\}} m_{\ell}^t$.
\item With probability 
$\alpha$, reallocate the measurement from target $j$ to $\ell$
\begin{itemize}
\item if $\exists j \in \mathcal{T}_i$ such that $m_{j}^t > \left \lceil{\frac{\vert\mathcal{N}\vert \cdot m}{\vert\mathcal{T}\vert}}\right \rceil$ and the measurement for $\ell$ is the most accurate one of those satisfying $\mathrm{ arg min}_{q \in \{\mathcal{T}/\mathcal{T}_i\}} m_{q}^t$, or
\item if $m_j^t- m_{\ell}^t=1$, where $ m_j^t=\mathrm{max}_{q \in\mathcal{T}_i} m_q^t$ and $m_{\ell}^t=\mathrm{min}_{q \in \{\mathcal{T}/\mathcal{T}_i\}} m_q^t$, and if measurement for $\ell$ is more accurate than the one for $j$. 
\end{itemize} 
\item Transmit/receive measurements, and $\forall j \in\mathcal{T}$, execute~\eqref{eq_predict_1}-\eqref{eq_predict_2} and employ all available measurements in~\eqref{eq_update_1}-\eqref{eq_update_3}.
\end{itemize}
To account for time-varying accuracy measures, e.g., range and/or azimuth variances, or to deal with different model dynamics, 
algorithm can be reinitialized every $K$ time instants so as to search for other NE during the tracking process.

%

\section{Simulations}

In this section, we will demonstrate the performance of the proposed algorithm for track selection in multi-target tracking.



We consider an MFR network of $\vert \mathcal{N}\vert=3$  radars, each of them making $m=2$  measurements per scan and aiming at tracking $\vert\mathcal{T}\vert=5$ targets. The coordinates of radars are 
%
%
%
$(x_1,y_1)=[-10\,km$, $0\,km]$,  
$(x_2,y_2)=[3\,km$, $0\,km]$ and   
$(x_3,y_3)=[10\,km$, $0\,km]$.
The targets follow white noise constant velocity trajectories with initial  $x$, $y$-coordinates and velocities 
%
$\mathrm{x}_{1,0}= [1 \,km,$  $6\,km$, $0.5 \,km/s$, $0.1 \,km/s]^T$, 
$\mathrm{x}_{2,0}= [0.5\,km,$ $7\,km,$ $0.35\,km/s,$ $-0.1\,km/s]^T$,
$\mathrm{x}_{3,0}= [1.5\,km,$ $3.0\,km,$ $-0.3\,km/s,$ $ 0\,km/s]^T$,
$\mathrm{x}_{4,0}= [2.0\,km,$ $ 4.0\,km,$ $ -0.2\,km/s,$ $ 0.1\,km/s]^T$ and
$\mathrm{x}_{5,0}= [2.5\,km,$ $ 5.0\,km,$ $ 0.3\,km/s,$ $ 0.2\,km/s]^T$. 
Initial guesses $\mathrm{x}_{j,0\vert0}$ are noisy versions of the initial states $\mathrm{x}_{j,0}$ and initial covariances are equal to $P_{j,0\vert0}=P_{0\vert0}=\mathrm{diag}\big\{(0.1\,km)^2,\break (0.1\,km)^2, (0.1\,km/s)^2, (0.1\,km/s)^2\big\}$. 
%
%
The update time is $t_u=0.25\,s$, and in order to model moderate maneuverability,  $\sigma^2_w$ is set to $2.5 \cdot 10^{-5} \,km^2/{s^3}$. 
%
Also, the standard deviation in azimuth is $\sigma^{(i)}_{a_j}=\sigma_a=2 \,mrad$, while the range accuracy varies among the radars and targets as $\sigma^{(i)}_{r_j}=b_{i,j}\cdot \sigma_r$, where $\sigma_r=15\,m$ and $b_{i,j}$ is taken from the interval $[1, 4.5]$.
\begin{figure}[t]
	\begin{minipage}[b]{1.0\linewidth}
		\centering
		\centerline{\includegraphics[width=7.7cm]{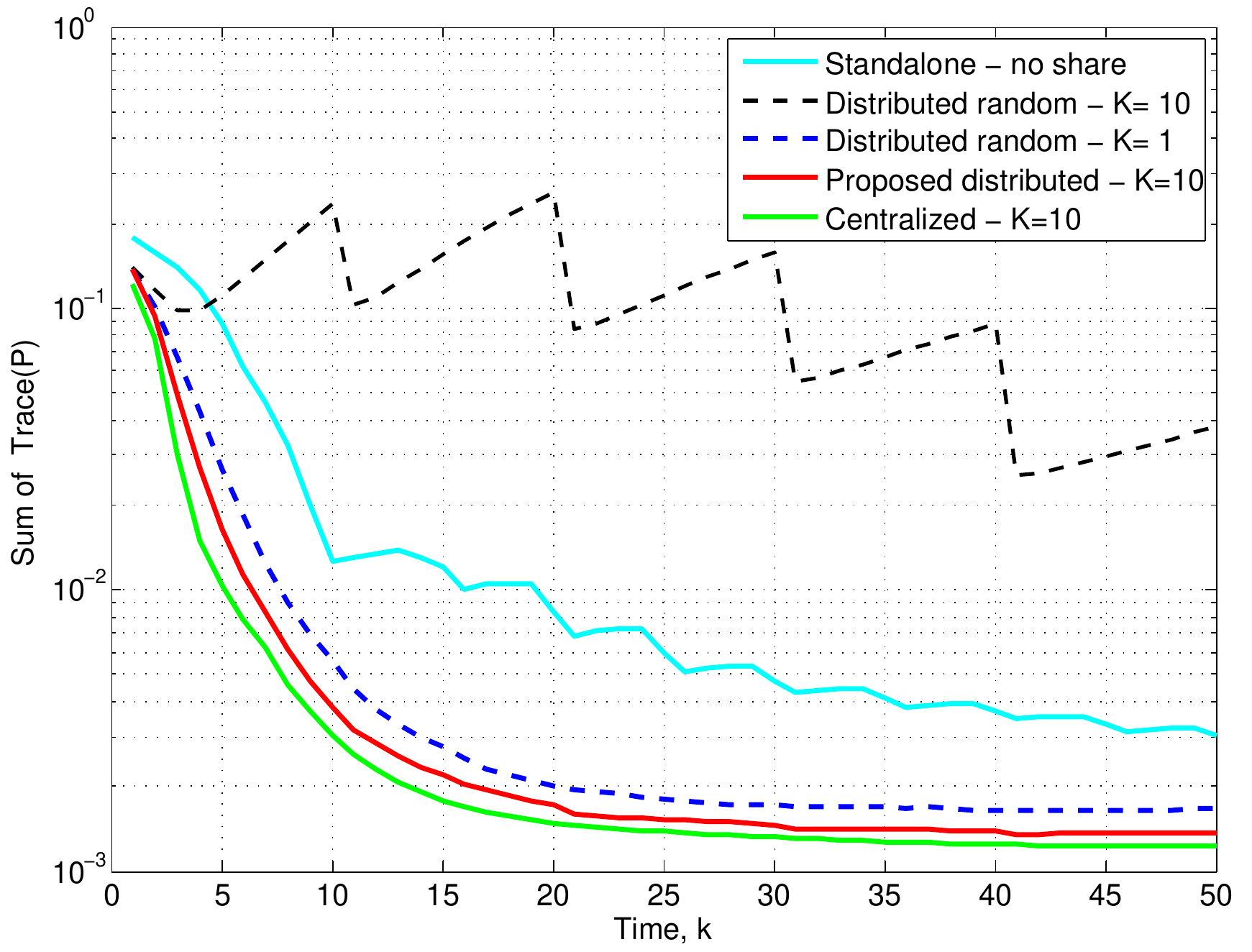}}
		\vspace{-0.10cm}
		\caption{Sum of traces of error covariance matrices for all targets during time.}
		\label{fig:Results_for_P}
	\end{minipage}
\end{figure}
%
Next, a comparison is made among the following strategies:
	(a) the standalone radar that does not send/receive measurements, and sequentially chooses $m=2$ different targets at each time instant;
	(b) distributed strategy where the radars exchange the measurements while each of them randomly changes its selection 
at each $K=10$ time instants;
	(c) same as in (b), except that 
 targets are being randomly chosen at each time instant;
	(d) the proposed distributed algorithm seeking NE for $\alpha=0.4$ while being reinitialized with $K=10$; and
	(e) an approximated centralized approach based on exhaustive search for optimal measurements-to-target allocation every $K=10$ time instants.

The results in Fig.~\ref{fig:Results_for_P} are averaged over 100 
realizations. 
Not surprisingly, due to the high process' dynamics, a standalone, non-cooperative radar experiences weak performance since it utilizes only its own measurements which are not sufficient to cover all targets. However, although approach in (b) uses 3x2=6 measurements, due to the lack of coordination it performs poorly. Note that  the distributed random strategy can be significantly improved if strategies are constantly being changed, given that there are no track migration costs involved. However, the proposed distributed algorithm that learns underlying NE outperforms aforementioned strategies and closely approaches to the performance of the centralized one while 
being much more efficient in terms of complexity.

\section{Conclusions}

We have formulated a track selection problem for multi-target tracking in a network of MFR radars. The problem has been tackled using the non-cooperative game theory. 
The  Nash equilibria of the underlying anti-coordination games have been analyzed and 
a simple yet effective distributed algorithm %
that looks for the equilibria points 
has been proposed.
It introduces a balancing effect in the track selection 
which makes it be particularly convenient for the settings with high dynamics. %
Also, %
it closely approximates the centralized performance while mitigating its inherent complexity. Our future work may consider extending the results for different communication topologies and for cases where not all radars have the same interests.

\clearpage

\bibliographystyle{IEEEbib}
\ninept
\bibliography{Nikola2_radar_conf}

\begin{thebibliography}{10}

\bibitem{hero2011sensor}
A.~O. Hero and D.~Cochran,
\newblock ``Sensor management: {P}ast, present, and future,''
\newblock {\em Sensors Journal, IEEE}, vol. 11, no. 12, pp. 3064--3075, 2011.

\bibitem{sabatini_tarantino_1994}
S.~Sabatini and M.~Tarantino,
\newblock {\em Multifunction array radar},
\newblock Artech House, 1994.

\bibitem{richards_scheer_holm_2010}
M.~A. Richards, J.~Scheer, and W.~A. Holm,
\newblock {\em Principles of modern radar},
\newblock SciTech Publishing, 2010.

\bibitem{blackman1999design}
S.~S. Blackman and R.~Popoli,
\newblock {\em Design and Analysis of Modern Tracking Systems},
\newblock Artech House radar library. Artech House, 1999.

\bibitem{Waveform_agile_4775880}
S.~P. Sira, Y.~Li, A.~Papandreou-Suppappola, D.~Morrell, D.~Cochran, and
  M.~Rangaswamy,
\newblock ``Waveform-agile sensing for tracking,''
\newblock {\em IEEE Signal Processing Magazine}, vol. 26, no. 1, pp. 53--64,
  Jan 2009.

\bibitem{Ding_4564804}
Z.~Ding,
\newblock ``A survey of radar resource management algorithms,''
\newblock in {\em Proc. Canadian Conference on Electrical and Computer
  Engineering}, May 2008, pp. 001559--001564.

\bibitem{katsilieris2015sensor}
F.~Katsilieris,
\newblock {\em Sensor management for surveillance and tracking: An operational
  perspective},
\newblock Ph.D. thesis, TU Delft, Delft University of Technology, 2015.

\bibitem{charlish2011autonomous}
A.~Charlish,
\newblock {\em Autonomous agents for multi-function radar resource management},
\newblock Ph.D. thesis, UCL (University College London), 2011.

\bibitem{narykov2013algorithm}
A.~S. Narykov, O.~A. Krasnov, and A.~Yarovoy,
\newblock ``Algorithm for resource management of multiple phased array radars
  for target tracking,''
\newblock in {\em Proc. 16th Int. Conference on Information Fusion (FUSION
  2013)}. IEEE, 2013, pp. 1258--1264.

\bibitem{van1993phased}
G.~van Keuk and S.~S. Blackman,
\newblock ``On phased-array radar tracking and parameter control,''
\newblock {\em IEEE Transactions on Aerospace Electronic Systems}, vol. 29, pp.
  186--194, 1993.

\bibitem{koch1999adaptive}
W.~Koch,
\newblock ``On adaptive parameter control for phased-array tracking,''
\newblock in {\em Signal and Data Processing of Small Targets}, 1999, pp.
  444--455.

\bibitem{coetzee2005multifunction}
S.~Coetzee, K.~Woodbridge, and C.~Baker,
\newblock ``Multifunction radar resource management using tracking
  optimisation,''
\newblock Tech. {R}ep., DTIC Document, 2005.

\bibitem{Hans_1591902}
J.~H. Zwaga and H.~Driessen,
\newblock ``Tracking performance constrained mfr parameter control: applying
  constraints on prediction accuracy,''
\newblock in {\em Proc. 8th International Conference on Information Fusion,
  2005}, July 2005, vol.~1, pp. 546--551.

\bibitem{hansen2006resource}
J.~Hansen, R.~Rajkumar, J.~Lehoczky, and S.~Ghosh,
\newblock ``Resource management for radar tracking,''
\newblock in {\em IEEE Conference on Radar, 2006}. IEEE, 2006, p.~8.

\bibitem{Djonin_4915772}
V.~Krishnamurthy and D.~V. Djonin,
\newblock ``Optimal threshold policies for multivariate {POMDP}s in radar
  resource management,''
\newblock {\em IEEE Transactions on Signal Processing}, vol. 57, no. 10, pp.
  3954--3969, Oct 2009.

\bibitem{shoham2008multiagent}
Y.~Shoham and K.~Leyton-Brown,
\newblock {\em Multiagent Systems: Algorithmic, Game-Theoretic, and Logical
  Foundations},
\newblock Cambridge University Press, 2008.

\bibitem{Han_2012_book}
Z.~Han, D.~Niyato, W.~Saad, T.~Baar, and A.~Hj{\o}rungnes,
\newblock {\em Game Theory in Wireless and Communication Networks: Theory,
  Models, and Applications},
\newblock Cambridge University Press, New York, NY, USA, 1st edition, 2012.

\bibitem{bacci_2015game}
G.~Bacci, S.~Lasaulce, W.~Saad, and L.~Sanguinetti,
\newblock ``Game theory for signal processing in networks,''
\newblock {\em arXiv preprint arXiv:1506.00982}, 2015.

\bibitem{Saraydar_983324}
C.~U. Saraydar, N.~B. Mandayam, and D.~Goodman,
\newblock ``Efficient power control via pricing in wireless data networks,''
\newblock {\em IEEE Transactions on Communications}, vol. 50, no. 2, pp.
  291--303, Feb 2002.

\bibitem{Larsson_4604732}
E.~G. Larsson and E.~A. Jorswieck,
\newblock ``Competition versus cooperation on the {MISO} interference
  channel,''
\newblock {\em IEEE Journal on Selected Areas in Communications}, vol. 26, no.
  7, pp. 1059--1069, September 2008.

\bibitem{scutari2008competitive}
G.~Scutari, D.~P. Palomar, and S.~Barbarossa,
\newblock ``Competitive design of multiuser mimo systems based on game theory:
  A unified view,''
\newblock {\em IEEE Journal on Selected Areas in Communications}, vol. 26, no.
  7, pp. 1089--1103, 2008.

\bibitem{ch1_j_Feleg4907463}
M.~Felegyhazi, M.~Cagalj, and J.-P. Hubaux,
\newblock ``Efficient mac in cognitive radio systems: A game-theoretic
  approach,''
\newblock {\em IEEE Transactions on Wireless Communications}, vol. 8, no. 4,
  pp. 1984--1995, April 2009.

\bibitem{Yu2013a}
C.~Yu, M.~van~der Schaar, and A.H. Sayed,
\newblock ``{Reputation design for adaptive networks with selfish agents},''
\newblock {\em Proc. 2013 IEEE 14th Workshop on Signal Processing Advances in
  Wireless Communications (SPAWC)}, pp. 160--164, June 2013.

\bibitem{GT_icassp2015}
N.~Bogdanovi\'{c}, D.~Ampeliotis, and K.~Berberidis,
\newblock ``Coalitional game theoretic approach to distributed adaptive
  parameter estimation,''
\newblock in {\em Proc. 2015 IEEE Int. Conference on Acoustics, Speech and
  Signal Processing (ICASSP)}, April 2015.

\bibitem{g_2012_game}
S.~Gogineni and A.~Nehorai,
\newblock ``Game theoretic design for polarimetric {MIMO} radar target
  detection,''
\newblock {\em Signal Processing}, vol. 92, no. 5, pp. 1281--1289, 2012.

\bibitem{code_design_piezzo2013non}
M.~Piezzo, A.~Aubry, S.~Buzzi, A.~De~Maio, and A.~Farina,
\newblock ``Non-cooperative code design in radar networks: a game-theoretic
  approach,''
\newblock {\em EURASIP Journal on Advances in Signal Processing}, vol. 2013,
  no. 1, pp. 1--17, 2013.

\bibitem{Jammer_6025317}
X.~Song, P.~Willett, S.~Zhou, and P.~B. Luh,
\newblock ``The {MIMO} radar and jammer games,''
\newblock {\em IEEE Transactions on Signal Processing}, vol. 60, no. 2, pp.
  687--699, Feb 2012.

\bibitem{Bacci_6250454}
G.~Bacci, L.~Sanguinetti, M.S. Greco, and M.~Luise,
\newblock ``A game-theoretic approach for energy-efficient detection in radar
  sensor networks,''
\newblock in {\em Proc. 7th Sensor Array and Multichannel Signal Processing
  Workshop (SAM)}, June 2012, pp. 157--160.

\bibitem{Chen_coop_game_7104065}
H.~Chen, S.~Ta, and B.~Sun,
\newblock ``Cooperative game approach to power allocation for target tracking
  in distributed {MIMO} radar sensor networks,''
\newblock {\em IEEE Sensors Journal}, vol. PP, no. 99, pp. 1--1, 2015.

\bibitem{charlish2012multi}
A.~Charlish, K.~Woodbridge, and H.~Griffiths,
\newblock ``Multi-target tracking control using continuous double auction
  parameter selection,''
\newblock in {\em Proc. 15th International Conference on Information Fusion
  (FUSION 2012)}. IEEE, 2012, pp. 1269--1276.

\bibitem{nisan2007algorithmic}
N.~Nisan, T.~Roughgarden, E.~Tardos, and V.~V. Vazirani,
\newblock {\em Algorithmic Game Theory},
\newblock Cambridge University Press, 2007.

\bibitem{bramoulle2007anti}
Y.~Bramoull{\'e},
\newblock ``Anti-coordination and social interactions,''
\newblock {\em Games and Economic Behavior}, vol. 58, no. 1, pp. 30--49, 2007.

\bibitem{book_rasmusen_2001}
E.~Rasmusen,
\newblock {\em Games and information},
\newblock Blackwell Publishers, 2001.

\end{thebibliography}

\end{document}